\documentclass{emulateapj}
\usepackage{apjfonts}

\newcommand{\timav}{\langle \dot M \rangle}
\newcommand{\be}{\begin{eqnarray}}
\newcommand{\ee}{\end{eqnarray}}
\shorttitle{Seismology of the WD in GW Lib} 

\begin{document}

\submitted{Accepted by ApJL}
\title{ Seismology of the Accreting White Dwarf in GW Lib } 

\author{Dean M. Townsley\altaffilmark{1},
Phil Arras\altaffilmark{2},
and Lars Bildsten\altaffilmark{1,2}}

\altaffiltext{1}{Department of Physics, Broida Hall,
University of California, Santa Barbara, CA 93106; townsley@physics.ucsb.edu}
\altaffiltext{2}{Kavli Institute for Theoretical Physics, Kohn Hall,
University of California, Santa Barbara, CA 93106; arras@kitp.ucsb.edu,
bildsten@kitp.ucsb.edu}

\begin{abstract}

We present a first analysis of the g-mode oscillation spectrum for
the white dwarf (WD) primary of GW Lib, a faint cataclysmic variable
(CV). Stable periodicities have been observed from this WD for a number
of years, but their interpretation as stellar pulsations has been
hampered by a lack of theoretical models appropriate to an accreting
WD. Using the results of Townsley and Bildsten, we construct accreting
models for the observed effective temperature and approximate mass of
the WD in GW Lib. We compute g-mode frequencies for a range of accreted
layer masses, $M_{\rm acc}$, and long term accretion rates, $\timav$.
If we assume that the observed oscillations are from $\ell=1$ g-modes,
then the observed periods are matched when $M\approx 1.02M_\odot$,
$M_{\rm acc}\approx 0.31\times10^{-4}M_\odot$ and $\timav\approx 7.3\times
10^{-11}M_\odot$ yr$^{-1}$. Much more sensitive observations are needed
to discover more modes, after which we will be able to more accurately
measure these parameters and constrain or measure the WD's rotation rate.

\end{abstract}


\defcitealias{TownBild03}{TB}

\section{Introduction}

Dwarf Novae (DN) are the subset of CVs with low time-averaged accretion rates
$\timav\lesssim10^{-9}M_\odot \ {\rm yr}^{-1}$ and thermally unstable accretion
disks that lead to sudden accretion events which interrupt the otherwise
quiescent state. In quiescence,
the UV (and sometimes optical) emission from the binary is dominated by
light from the WD surface, allowing for a measurement of the WD's $T_{\rm
eff}$. These $T_{\rm eff}$'s are much hotter than expected for a WD of the age
of the binary (a few Gyr) and 
must be related to accretion \citep{Sion99}.  Calculations of the 
heating of the deep interior of the WD by the prolonged accretion
\citep{TownBild04}
explains the observed values of $T_{\rm eff}$ and yields a unique
relationship between
$\timav$, the WD mass, $M$, and the orbital period $P_{\rm orb}$
\citep{TownBild03}.

GW Lib is one of the shortest orbital period CVs known, $P_{\rm orb} = 77$
min \citep{Thoretal02}, and therefore has a very low $\timav\sim 5\times
10^{-11} M_\odot$ yr$^{-1}$ \citep{TownBild03}.  Only one
disk outburst has been observed from GW Lib
\citep{Gonz83}.  Much later photometric observations during quiescence led to
the discovery of
periodic variability \citep{vanZetal00} similar to that of
isolated WDs which pulsate due to non-radial g-modes
(see \citealt{Brad98} for a review of the DAV WDs).  
The highest signal to noise photometric observations comprise two weeks of
single-site data taken in 1998 with some supplement from other longitudes
\citep{vanZetal04}.  There are three clear periodicities, listed in Table
\ref{tab:periods}, with some evidence for mild period variability.

\begin{deluxetable}{cc}
\tablewidth{0pt}
\tablecaption{\label{tab:periods}Observed Periods}
\tablehead{ System & Principal Period, $P$\\ & (seconds) }
\startdata
GW Lib & $236$, $377$, $646$\\
SDSS 1610 & 345, 607\\
SDSS 0131 & $\simeq$ 330, $\simeq$ 600\\
SDSS 2205 & $\simeq$ 330, $\simeq$ 600
\enddata
\end{deluxetable}

Three additional CV WD pulsators have been found
\citep{WarnWoud03,WoudWarn04} by taking
photometric time series of DN identified in the Sloan Digital Sky
Survey (SDSS; \citealt{Szkoetal03}) which show WD spectral features in the
optical.  These objects and their dominant periods
are also listed in Table \ref{tab:periods}.  At this discovery
fraction we expect
$\simeq$15 CV WD pulsators total by the end of the SDSS.
Seismology of the CV WD pulsators provides the prospect of well-determined
masses for these systems and the parameters we derive for GW Lib are
the first step in this process. 

Mass determinations are especially interesting due to possible implications
for progenitor systems of Type Ia supernovae.  Thought to be accreting WDs near
the Chandrasekhar mass \citep{HillNiem00}, these progenitors are closely
related to the CV population, but the precise nature of this relationship is
unknown.  Important clues in this mystery lie in how the masses of CV primary
WDs change over the accreting lifetime of the binary, but progress is
hampered by the difficulty of measuring CV primary masses \citep{Patt98}.
This mass evolution, as well as probing the $M_{\rm acc}$ directly with
seismology, is also important for determining how much, if any, of the
original WD material has been ejected into the ISM in classical nova
outbursts, contributing to the ISM metallicity \citep{Gehretal98}.

We begin in \S 2 by discussing the observed properties of the WD in GW Lib
and constructing accreting WD models that are consistent with the
observations. Section 3 discusses the g-mode properties for the accreting
model of GW Lib, and in Section 4 we show what can be learned about the
accreting WD from the observed mode periods. We conclude in Section 5 with a
discussion of future work.


\section{Observed Properties of the White Dwarf in  GW Lib}

Though the periods make it clear that the oscillations in GW Lib are
non-radial g-modes, without secure identification of the radial and angular
quantum numbers ($n$ and $\ell$), it is difficult to carry out the
seismology.   Hence, we start by using the  binary's observed properties to
constrain a few of the WD's properties, thus reducing the range of
possibilities we need to consider in our modeling efforts. 
  
\subsection{Limits from Distance and $T_{\rm eff}$ }

\begin{figure}
\plotone{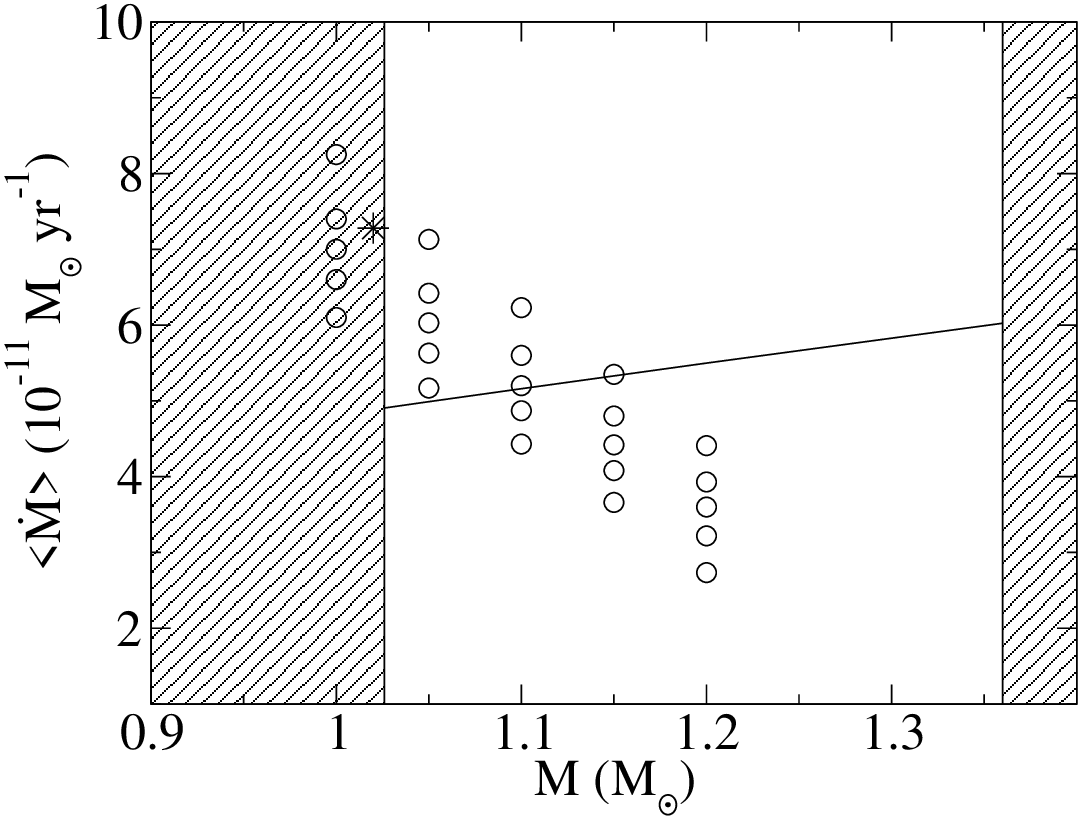}
\caption{\label{fig:constraints}
Parameter space constraints from non-seismological observations.
Distance, UV flux and UV spectral measurements constrain $M$ (diagonal
shading) and gravitational wave radiation provides a (mass dependent) lower
limit on $\timav$ (solid line).  Actual models considered here are indicated by
the circles.
The $*$ indicates our best fit to the observed pulsations.
}
\end{figure}

We  use three observations: (1) the parallax (2) a $T_{\rm eff}$-$\log
g$ relationship determined from the UV spectrum, and (3) the $P_{\rm orb}$.
The parallax of GW Lib has been measured from the ground \citep{Thor03} and
gives a distance of $104^{+30}_{-20}$ pc. Current  UV spectroscopic
observations can only constrain $T_{\rm eff}$ and $\log g$ to a
well-determined linear relationship (see \citealt{Howeetal02} for a
discussion).  In the case
of GW Lib, $T_{\rm eff}/{\rm K} = 14700+2000\log_{10} g_8$ where $g_8$
is the surface gravity measured in $10^8$ cm
s$^{-2}$ \citep{SzkoGWLib02}.  By using this, the UV flux measured from the
same observation, the distance, and a WD mass-radius relation
we find $M=1.03$ to $1.36M_\odot$ from the range of allowed
distances.  This constraint excludes the diagonally shaded regions  in Figure
\ref{fig:constraints}.

Another parameter needed is the time-averaged accretion
rate, $\timav$.  For each $M$, $P_{\rm orb}$ sets
the orbital separation, and with a mass-radius relation for the Roche Lobe
filling companion (we use \citeauthor{KolbBara99}'s 1999 for a low mass main sequence
star) this allows  a derivation of a mimimum $\timav$ due
only to gravitational radiation from the orbit.
This limit is shown by the solid line in Figure
\ref{fig:constraints}, and can be high if the system has
passed through the period minimum, so that the companion is a
sub-stellar object which is out of thermal equilibrium and has a
lower mean density.

Optical spectroscopy (\citealt*{Szkoetal00}; \citealt{Thoretal02})
has given lower $T_{\rm
eff}$ (11000 K and 13220 K respectively), however both of these studies used
gravities much lower than the allowed range derived above ($\log g=8$ and
7.4 as opposed to 8.6 for $M=1.0M_\odot$),
and were unable to effectively account for contamination by the quiescent
accretion disk, both of which would lead to low fitted $T_{\rm eff}$.
The UV spectrum is the most reliable indicator of
$T_{\rm eff}$.

\subsection{The Accreting White Dwarf Structure}

The WD interior structure is from  \citet[hereafter TB]{TownBild04}.  We
chose a simple compositional structure consisting of a solar composition
accreted layer of mass $M_{\rm acc}$ on the WD core, an equal mixture
of $^{12}$C and $^{16}$O. For calculation of the buoyancy properties, a
smooth transition region of $0.2$ times the local pressure scale height was
put in between these layers (see TB for a discussion of the diffusion
timescales).  With this compositional structure, a WD model is parameterized
by $M$, $T_{\rm core}$, $\timav$ and $M_{\rm acc}$, and compression of material
by accretion powers a surface luminosity  $L$.
TB related $T_{\rm core}$ to $\timav$ by finding the equilibrium
state where between classical novae outbursts (as $M_{\rm acc}$ is growing)
the WD core would suffer no net heating or cooling.  For a fixed $T_{\rm
eff}$ at the observed value, TB also relate $\timav$ and
$M_{\rm acc}$, with a higher $M_{\rm acc}$ implying a lower $\timav$ for the
same $T_{\rm eff}$.  Using these two constraints we are left with $M$ and
$M_{\rm acc}$ as free parameters, where we only consider $M_{\rm acc} \le
M_{\rm ign}\sim 10^{-4}M_\odot$, the classical nova ignition mass. 

This smaller grid of models are shown by the circles in Figure
\ref{fig:constraints}, and have $M/M_\odot=1.0,  1.05, 1.1, 1.15,$ and 1.20,
each with $M_{\rm acc} = 0.1$, 0.3, 0.5, 0.7, and 0.9$M_{\rm ign}$.  The
ignition masses for these models were $M_{\rm ign}= 1.40$, 1.32, 1.22, 1.14,
and $1.08\times 10^{-4}M_\odot$.  The model at $M=1.05M_\odot$ and $M_{\rm
acc}= 0.3M_{\rm ign} = 0.40\times 10^{-4}M_\odot$ has $T_{\rm eff}=16070$ K
(that implied for this $M$ by the UV spectrum), $\timav = 6.4\times
10^{-11}M_\odot$ yr$^{-1}$, and $T_{\rm core} = 5.9\times 10^6$ K.  The
downward trend in $\timav$ with increasing $M$ is due to $L$ decreasing with
the WD radius with $T_{\rm eff}$ constrained to the measured relation.

\section{ G-modes in GW Lib's White Dwarf}

In a nonrotating star, all variables can be decomposed into
spherical harmonics $Y_{lm}(\theta,\phi)$. We ignore the perturbed
gravitational potential (the Cowling approximation) and work in the adiabatic
approximation.
The linearized momentum and energy equations are
then written in terms of the radial displacement $\xi_r(r)$ and
the Eulerian pressure perturbation $\delta p(r)\equiv \rho \psi(r)$ as
(See e.g. eq.14.2 and 14.3 in \citealt{Unnoetal89})
\be
\frac{d\psi}{dr} &=& \frac{N^2}{g} \psi -(N^2-\omega^2)\xi_r\ ,
\nonumber \\
\frac{d\xi_r}{dr} &=&
-\left( \frac{1}{c^2} - \frac{l(l+1)}{r^2 \omega^2} \right) \psi
- \left( \frac{2}{r} - \frac{g}{c^2} \right) \xi_r\ .
\ee
Here $\omega=2\pi/P$ is the mode frequency, $P$ is the period,
$c=(\Gamma_1 p/\rho)^{1/2}$ is the adiabatic sound speed, $g=GM(r)/r^2$ is
the downward gravitational acceleration, and $N^2=-g \left(d\ln\rho/dr -
\Gamma_1^{-1} d\ln p/dr \right)$ is the square of the Brunt-V\"ais\"al\"a
frequency.  For boundary conditions, we impose zero Lagrangian pressure
perturbation, $\Delta p/\rho=\psi - g\xi_r=0$, at the top of the model
and $\xi_r=0$ at the interface between the solid core and surrounding
liquid.  This interface occurs at the freezing point for a Coulomb solid
(\citealt{BildCutl95}; \citealt{MontWing99}) and
g-modes cannot penetrate into the solid core as
their frequency is too low to excite significant shearing motion of
the Coulomb solid.
A solid core is present in the GW Lib model due to
the high mass.

\begin{figure}
\plotone{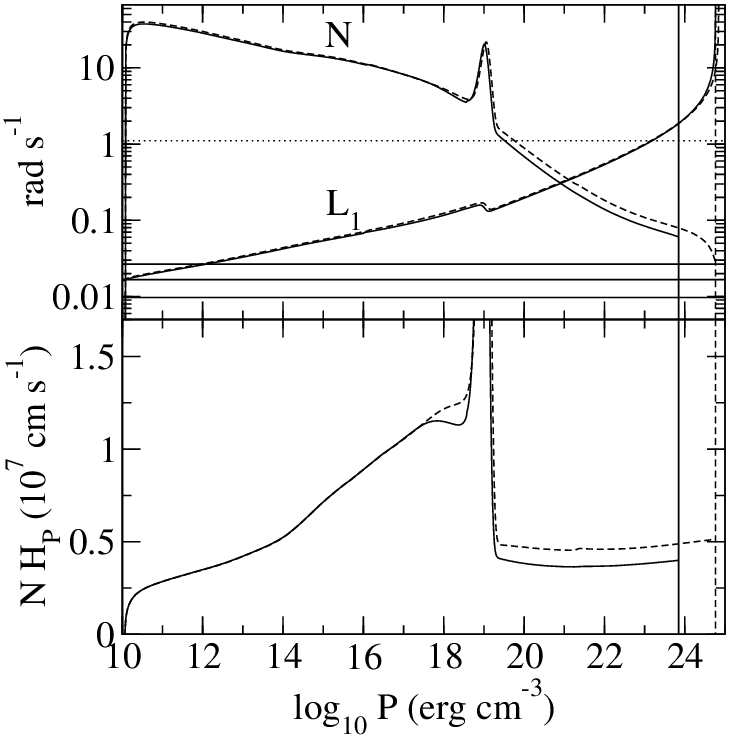}
\caption{ \label{fig:propint}
Comparison of propagation diagram (top panel) and WKB integrand
(bottom panel) for an accreting model (solid
lines) and
non-accreting model (dashed line) with the same composition and $T_{\rm eff}$.
The vertical solid (dashed) line on the right indicates the
liquid-solid boundary.
The solid horizontal lines are the observed frequencies and the dotted
horizontal line is the breakup frequency.
The accreting model has $M=1.05M_\odot$, $M_{\rm
acc}= 0.3M_{\rm ign} = 0.40\times 10^{-4}M_\odot$, $T_{\rm eff}=16070$ K
(that implied for this $M$ by the UV spectrum), $\timav = 6.4\times
10^{-11}M_\odot$ yr$^{-1}$, and $T_{\rm core} = 5.9\times 10^6$ K.  The
non-accretor has $T_{\rm core}=9.3\times10^6$ K.
}
\end{figure}

The propagation cavity for the observed short wavelength g-modes
(\citealt{Unnoetal89}) in GW Lib is bounded from above by $\omega^2
= L_l^2=l(l+1)c^2/r^2$, the Lamb frequency, and by
the solid core below.
The propagation diagram for the model with $M=1.05M_\odot$ and $M_{\rm
acc} = 0.3M_{\rm ign}=0.40\times10^{-4}M_\odot $ is shown
in the upper panel of Figure \ref{fig:propint}.
The large peak in $N$ at $\log_{10} p\simeq 19$ is due to the
change in mean molecular weight in the transition layer from the solar
composition accreted envelope to the C/O core.  In the roughly constant
flux envelope, $N^2 \simeq g/z$ where $z$ is the depth
from the surface. In the degenerate core, $N^2 \simeq (g/H_p)(k_bT/E_F)$,
where $E_F$ is the electron Fermi energy and
$H_p = p/\rho g$ is the pressure scale height.
In the propagation zone of the wave, the WKB dispersion
relation for low frequency g-modes gives a radial wavenumber
$k_r=(N/\omega)[l(l+1)]^{1/2}/r$.
The quantized WKB phase $\varphi$ accrued by the wave in a given region is
$\varphi = \int dr k_r = [l(l+1)]^{1/2} \omega^{-1} \int N dr/r$ so that $\omega
\propto \int NH_p\,d\ln p$.
The integrand is shown in the bottom panel
of Figure \ref{fig:propint}, representing the number of nodes per
decade in pressure.  Shown are two WD models: an accreting model with
$M=1.05M_\odot$ and $M_{\rm acc} = 0.3M_{\rm ign}=0.40\times 10^{-4}M_\odot$
used in our mode
analysis (solid line), and a non-accreting model with the same $T_{\rm
eff}$ (dashed line) and composition.

Not shown in Figure \ref{fig:propint} is the difference due to the envelope
mean molecular weight, $\mu$, from the case of a pure H or He envelope to one
of solar compositon.  Such a change is reflected in the periods as
$P_n\propto\mu^{0.5}$, since the envelope is well-approximated
by an $n=4$ polytrope.
The non-accreting model also has a higher $T_{\rm core}$
than the accreting model
by about 50\%,
leading to two important effects: (1) the WKB integrand has a higher
value in the core for the non-accreting model ($\omega \propto T_{\rm
core}^{1/2}$), and (2) the solid core is smaller,
pushing the inner boundary condition deeper into the star. Both effects
directly influence the observed periods, and period spacings, 
hence {\it it is essential to use a WD model including the effects of 
compressional heating and nuclear burning}, rather than a passively
cooling WD model.



\section{Inferring the WD Properties from the Periods} 

The existing optical time-series photometry of GW Lib \citep{vanZetal04}
consists of 7 time series taken in 1997, 1998, and 2001.  Here we focus on
the best of these, the two weeks of data from May 1998.  
Table \ref{tab:periods} represents our estimate of the three periods.
From the $O-C$ phase plots shown by \citet{vanZetal04}, the
periodicity near 646 s varies with $\dot\nu =\dot\omega/2\pi\simeq
- 10^{-11}$ Hz s$^{-1}$, and the phase is tracked by the
observations.  For that near 377 s, the phase is lost,
and therefore it is inconclusive whether
there are multiple closely spaced periods or just unresolved variability.
The highest frequency is quite stable, and though close to a sum is
unlikely to be a combination frequency.  Splittings $\simeq 1$ $\mu$Hz
are ubiquitous in the Fourier transforms, but their origin is unclear.  The
Doppler shift due to the orbit should have a magnitude of $0.5$ to 1.3
$\mu$Hz for the 646 and 236 s periods respectively. The frequency resolution
set by the inverse length of the time series is also $\sim \mu$Hz.

As mentioned by \citet{vanZetal04}, the mode frequencies might drift due to
the cooling of the material accreted in the last DN outburst.  The magnitude
and sign of this drift can be approximated using the cooling time of the freshly
accreted material and its fractional contribution to the WKB integral
discussed above.  For a recurrence time of 20 years (the time since the last
outburst) with $\timav\simeq 5\times 10^{-11}M_\odot$ yr$^{-1}$,
$\Delta M\simeq10^{-9}M_\odot$ was deposited in the last outburst.
For $M=1.0M_\odot$ the base of this layer has $p_{b} \simeq
2\times 10^{14}$ dyne cm$^{-2}$,  and a thermal time of
$\tau_{\rm th} = c_p T_b \Delta M / L \sim 1$ yr $(p_b/10^{14}{\rm dyne\
cm^{-2}})^{1.2}$, implying that this layer relaxes to close to the static
solution in a few years after the outburst, though a small
amount of cooling continues.  As can be seen from the lower panel in Figure
\ref{fig:propint}, the mode frequencies are largely determined deeper in
the WD;  however, this outer layer does have a contribution $\int N dz\propto
p_b^{1/10}\propto T_b^{1/2}$, for an
$n=4$ polytrope.  This accounts for about 20\% of the whole integral, and
leads to $\dot \nu \simeq0.2\nu\dot T/2T = -\nu/10\tau_{\rm th} \sim
-10^{-12}{\rm Hz\ s^{-1}}$ for $\nu = 1/646$ s, close enough to the observed
drift to warrant further calculations of this effect, which we do not
undertake here.
Shorter period (lower $n$) modes reside deeper in the star, making them
less vulnerable to transient heating and cooling effects.

\begin{figure}
\plotone{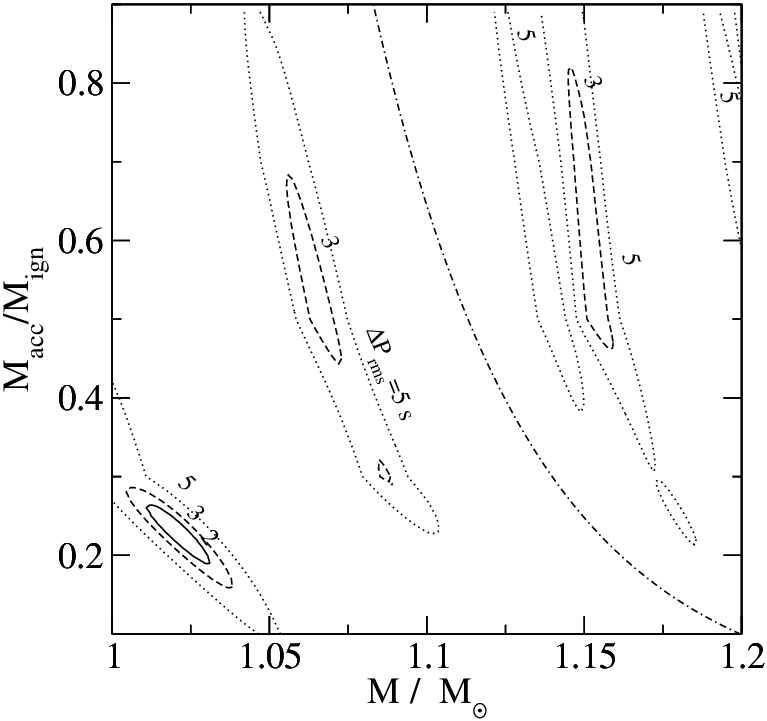}
\caption{
Contours in the root mean squared difference between the mode periods
observed and the closest modes in the model, $\Delta P_{\rm rms}$.  The best
solution lies in the lower left hand corner where $M=1.02M_\odot$, $M_{\rm
acc} = 0.23 M_{\rm ign} = 0.31\times 10^{-4}M_\odot$.  The dot-dashed line
demonstrates approximately how $M_{\rm acc}/M_{\rm ign}$ scales with $M$ for
a fixed period and $n$.
\label{fig:rms_cont}
}
\end{figure}

We have calculated the g-mode spectrum of model WDs with parameters
on the grid indicated in Figure \ref{fig:constraints} for $\ell=1$.
By interpolating within this grid we compute the model periods
$P_{n}(M,M_{\rm acc})$.  The root mean square period difference between
the observed three modes and the models is defined as $\Delta P_{\rm
rms}^2=\sum_{i=1}^3 {\rm min}_{\{n\}}[(P_{n} - P_i)^2]/3$, where the
observed periods have been indexed with $i=1,2,3$. The contours of $\Delta
P_{\rm rms}$ are shown in Figure \ref{fig:rms_cont}, implying a best fit
solution with $\Delta P_{\rm rms} = 1.8$s for $M=1.02M_\odot$ and $M_{\rm
acc} = 0.23M_{\rm ign} = 0.31\times 10^{-4}M_\odot$. The mode identifications
for this best fit model are $n=3$, 8, and 17
for the three observed periods.  These do not appear to correspond to any
particular mode trapping pattern reflected in the mode kinetic energy.
For this model, we can predict additional
mode periods yet to be observed; up to $n=17$ these are 141, 191, 234,
268, 289, 310, 351, 377, 400, 432, 463, 492, 519, 553, 586, 617, 647, 678 s.

The contours for $\Delta P_{\rm rms}$ do not close around a single
solution, implying additional data is needed to get better constraints.
There are two other areas in the $M$-$M_{\rm acc}$ plane which also
provide fairly good matches ($P_{\rm rms}< 3$ s) to the observed periods.
These are centered near $(M/M_\odot,\ M_{\rm acc}/M_{\rm ign})$ of (1.06,
0.53) and (1.15, 0.6), and extend in the $M_{\rm
acc}$ direction.  Each of the three minima represent a different set
of mode IDs, the shallower ones having $n=4$, 9, 19 and 5, 11, 22 for
the 236, 377, and 646 s modes. The $M$, $M_{\rm acc}$ degeneracy can be
simply explained. For a mode trapped in the envelope, $P_n\propto n /
\int_{\rm env} N dz \propto n M_{\rm acc}^{-0.1}M^{-2.3}$ for a polytrope
of index $4$ and an assumed scaling $R\propto M^{-1.8}$, suitable for
$M=1.0$ to $1.2M_\odot$.  Thus fixing $n$ and $P_n$ defines the relation
$M_{\rm acc} /M_{\rm ign}\propto M^{-21.5}$, where $M_{\rm ign}\propto
M^{-1.5}$, leading to the degeneracy shown by the dot-dashed line
in Figure \ref{fig:rms_cont}.


\section{Conclusions and Uncertainties}
\label{sec:rotation}

We have found non-rotating accreting WD models which fit the
periodic variability observed in GW Lib, implying $M=1.02M_\odot$,
$M_{\rm acc}=0.31\times10^{-4}M_\odot$ and $\timav=7.3\times
10^{-11}M_\odot$ yr$^{-1}$.  This model is quite simple, but since it is
capable of accounting for the observed periods, a more complex model is
not called for at this time. A number of important issues were not dealt
with, such as the presence of a residual He layer due to inefficient
nova mass ejection, or our treatment of the accreted
layer-C/O boundary.  More detailed non-adiabatic calculations will be
undertaken to address mode excitation.
The most essential omission in our calculations is
the WD rotation, since it is expected to be spun up by accretion.  If the
WD is rapidly rotating, the model parameters derived here are likely to
be inaccurate.  Better observations are necessary for this next essential
parameter to be constrained.  These observations should accomplish two
goals (1) to clearly characterize the substructure or variability of the
377 s and 646 s periods, (2) probe for lower amplitude signals.  A larger
number of identified modes are essential for conclusive seismology.

Rotation may significantly modify the mode frequencies. Since the maximum
rotation rate for our GW Lib models is $\sim 1 {\rm\
rad\ s^{-1}}$ and the observed mode frequencies are $\omega\sim 10^{-2} {\rm
\ rad\ s^{-1}}$, {\it spin frequencies above 1\% breakup will cause large
frequency shifts for observed modes.}  In fact \citet{SzkoGWLib02}
estimate $v\sin i< 300\ {\rm km \ s^{-1}}$, allowing spin rates up to 6\% of
breakup even for high $i$.  Rotation rates this large lead to a qualitative
change in mode properties.  When the spin frequency $\Omega_{\rm
spin} \lesssim 0.5\omega$, rotation can be treated as a small perturbation.
Also, observationally this rotational frequency splitting allows a determination
of the $l$ quantum number.  If, however, $\Omega_{\rm spin} \gtrsim
0.5\omega$, the Coriolis force has a ``nonperturbative'' effect on the mode
frequencies, modifying the relation between the frequency and quantum numbers
\citep*{Bildetal96}.  Hence accreting WDs may provide a unique proving ground
for techniques of seismology for rapidly rotating stars.


This work was supported by the National
Science Foundation under grants PHY99-07949, AST02-05956, and 0201636, and by
NASA through grant AR-09517.01-A from STScI, which is operated by
AURA, Inc., under NASA contract NAS5-26555.
Phil Arras is a NSF AAPF fellow. 


\bibliography{gwlib}

\end{document}